\documentstyle[12pt,citesort]{article}

\def \CV {NS5$_{S^2}$ }
\def \CVD {D5$_{S^2}$ }
\def \ae {{\rm E}} 
\def \Epsilon {{\cal E}} 

\catcode`\@=11

\def\bea{\begin{eqnarray}}
\def\eea{\end{eqnarray}}
\def\beann{\begin{eqnarray*}}
\def\eeann{\end{eqnarray*}}
\def\beq{\begin{equation}}
\def\eeq{\end{equation}}
\def\ba{\begin{array}}
\def\ea{\end{array}}
\def\ben{\begin{enumerate}}
\def\een{\end{enumerate}}
 \def \l {\lambda}
 \def\m {\mu}
 \def \la {\label}
 \def\be{\begin{equation}}
\def\ee{\end{equation}}
 \def \zz{{\bf z}}

\renewcommand{\thefootnote}{\fnsymbol{footnote}}

\newcommand{\newsection}{    
\setcounter{equation}{0}
\section}
\def\appendix#1{
  \addtocounter{section}{1}
  \setcounter{equation}{0}
  \renewcommand{\thesection}{\Alph{section}}
  \section*{Appendix \thesection\protect\indent \parbox[t]{11.15cm}
  {#1} }
  \addcontentsline{toc}{section}{Appendix \thesection\ \ \ #1}
  }

\def \ci {\cite}
\newcommand{\rf}[1]{(\ref{#1})}
\def \la {\label}
\def \const {{\rm const}}

\renewcommand{\thesection}{\arabic{section}}
\def \ha{{\textstyle { 1 \ov 2}} }

\def \r {\rho} 
\def \fo{{ 1 \ov 4}}

\font\mybb=msbm10 at 11pt

\def\bb#1{\hbox{\mybb#1}}

\def\bZ {\bb{Z}}
\def\bR {\bb{R}}

\def\bC {\bb{C}}

\def \foot {\footnote}
 
\def \ov {\over}
\def \ha { { 1\ov 2}}
\def \we { \wedge}
\def \P { \Phi} \def\ep {\epsilon}

\def \ba {{B^2 \ov A^2}}
\def \tv   {{1 \ov 12}}
\def \go { g_1}\def \gd { g_2}\def \gt { g_3}\def \gc { g_4}\def \gp { g_5}
\def \F {{\cal F}}
\def \del { \partial}
\def \t {\theta}
\def \p {\phi}
\def \ee {\epsilon}
\def \te {\tilde \epsilon}
\def \ps {\psi}
\def \td {\tilde}
 
\def \bi{\bibitem}

\date{}

\begin{document}


\begin{titlepage}

\begin{center}
\hfill OHSTPY-HEP-T-00-30\\

\vskip  1 cm
\vskip 1 cm


%
{\Large \bf  Complex geometry of conifolds}
  \vskip 0.3 cm
{\Large \bf and 5-brane  wrapped on 2-sphere}
\vskip  1 cm

{\large G. Papadopoulos$^a$ and A.A. Tseytlin$^{b,}$\footnote{Also at
 Imperial College, London and Lebedev
Institute, Moscow.}}\\

\end{center}

\centerline{\it ${}^a$ Department of Mathematics}
\centerline{\it King's College London}
\centerline{\it  London WC2R 2LS, U.K. }

\vskip 0.4 cm

\centerline{\it ${}^b$ Department of Physics, The Ohio State
University,}
\centerline{\it Columbus, OH 43210-1106, USA}

\vskip 1.5 cm

\begin{abstract}
We  investigate solutions of type II supergravity 
which have the product $\bR^{1,3}\times M^6$ 
structure  with non-compact  $M^6$ factor  and 
which preserve at least four supersymmetries.
 In particular, we consider 
various  conifolds and the $N=1$ supersymmetric 
 ``NS5-brane wrapped on 2-sphere"
 solution recently  discussed in hep-th/0008001.
  In all of these cases,  we  explicitly 
construct the complex structures, and the  K\"ahler 
 and parallel (3,0) forms of the corresponding $M^6$. 
 In addition, 
we  verify that  the above  solutions preserve respectively 
eight and  four supersymmetries 
of the underlying type II theory.
We  also  demonstrate that the ordinary  and  fractional D3-brane
(5-brane wrapped on 2-cycle) solutions 
on singular, resolved and
deformed  conifolds, and the  (S-dual of)  NS5-brane wrapped
on 2-sphere  can be obtained  as special 
cases from  a universal  ansatz for the supergravity  fields, 
i.e. from a single 1-d action governing their radial
 evolution. We show that 
like the 3-branes on  conifolds, 
the NS5-brane on 2-sphere   background  can be 
found  as a solution of first  order system following from 
a superpotential.

\end{abstract}

\end{titlepage}
\def \tx {\textstyle}

\newpage
\renewcommand{\thefootnote}{\arabic{footnote}}
\setcounter{footnote}{0}

\newsection{Introduction}

Supergravity solutions that preserve some
of the supersymmetry of underlying theory have found many 
applications in the exploration of perturbative and non-perturbative
properties of string theory. 
An important   
 example is  the AdS/CFT correspondence which asserts 
  that string theory on the $AdS_5\times S^5$ 
 background  
is related  to $N=4$ supersymmetric
Yang-Mills (SYM) theory in four dimensions \ci{ads}.
Recently,  attempts have been made to extend this 
correspondence 
to find  string theory duals of $N=1$
 supersymmetric  gauge theories (see, in particular, 
 \ci{kn,kt,kst,mn} and also \ci{pg,gub,pt,cp}).

 A remarkable observation made in 
 \ci{mn} is that a  non-trivial 
  supersymmetric solution 
  constructed in \ci{cv} may be interpreted 
  as describing a near-throat region of a
  large number of  NS5-branes
   wrapped on 2-sphere. In what follows we refer to this solution
   as   \lq\lq NS5-branes
   wrapped on 2-sphere" and denote
   it as \CV.  The  S-dual of  \CV
  represents D5-branes wrapped on $S^2$, and thus
 string theory on this background
  was conjectured to be dual to pure $N=1$ SYM theory in
  four dimensions.

One of the aims of the present paper is to 
investigate the geometrical properties,  and
 clarify furher the residual supersymmetry, 
of  the ten-dimensional  
NS$\otimes$NS 
background of \ci{cv,mn}. 
In addition, the holomorphic geometry of 
singular, deformed and resolved Calabi-Yau conifolds
 will be explored.

The solutions that we shall consider  belong
 to the  class of ten-dimensional NS$\otimes$NS  backgrounds 
 which are products $M^{10}=M^4\times M^6$
 of a four- and a six-dimensional space with possibly a
  warp factor multiplying 
 the metric of $M_4$.  In addition, they  
   (i) preserve at least four of the original supersymmetries, 
and (ii) have non-trivial  dilaton $ 
 \P$  and  Kalb-Ramond field strength $H$.\foot{Non-constant
  dilaton is of course expected to be related 
to running gauge coupling on gauge theory side.
Such backgrounds may be of interest in the context of the  
general framework suggested in   \ci{pol}.}

An important  simplification 
in  investigating  NS$\otimes$NS  backgrounds is that
some of the Killing spinor equations are parallel transport
equations of a connection with torsion
(see,  e.g.,  Section 2). Since for the applications mentioned
above $M^4=\bR^{1,3}$, the $M^6$ part of spacetime
  has properties similar to those of Calabi-Yau spaces.
In fact,  the Killing spinor equations impose strong
restrictions on the existence of such solutions  \ci{stro}.
In particular, it was shown  in \cite{ip}
that if one assumes that $M^6$ 
is compact and the dilaton is a globally
defined function on it,  then 
the only such spaces that preserve
at least four supersymmetries  are  the
Calabi-Yau spaces with 
  $H=0$ and $\P=\const$. 
  
To have supersymmetric solutions 
 with {\it running}  dilaton,   one has to consider 
 backgrounds  for which $M^6$ is {\it non-compact}. 
The solution of \ci{cv,mn}, i.e.   
  \CV background,
 provides an example. The metric on the 
 spacetime $M^{10}=\bR^{1,3} \times M^6$
 is the direct sum of the flat metric on  $\bR^{1,3}$
 and the \CV metric on  $M^6$.
It turns out that  this six-space  has many similarities
with non-compact Calabi-Yau manifolds
like,   for example,  the  conifold and its resolved 
and deformed  versions \ci{can} (see also \ci{min,oht}).
These conifolds appeared  as transverse spaces
 in the 
(ordinary and fractional) D3-brane supergravity   solutions
constructed in \ci{kw,kt,kst,pt}.

The  expressions for the metrics
of the spaces that we shall investigate  are known, but
it  is not so for their basic  geometrical properties   
as {\it  complex  six-dimensional} manifolds. In particular,   
 complex structures, and K\"ahler  
 and parallel (3,0)-forms (see Section 2 for the details)
have not been explicitly  determined. 
Both the complex structures and
the (3,0)-forms are important for a  study of 
  {\it string theory} 
in these backgrounds.  For example,
 complex structures are used to
construct the  ($N=2$) supersymmetry transformations for the string
world-sheet action, 
 while the (3,0)-forms are associated
with conservation laws.\foot{The latter,  however, can develop 
anomalies at the quantum level \cite{hpw}.}
Other applications of  the complex
structures and parallel (3,0)-forms 
are  in the context of geometry and,  in
particular,  in the investigation of 
calibrated submanifolds. These,  in turn, 
have applications in the context of branes.

Let us  summarize the contents of this paper. 
We  shall explore below the complex-geometric 
properties of singular, deformed and resolved conifolds
 as well as of \CV space, 
 presenting   their
complex structures,  K\"ahler forms and parallel  (3,0)-forms of these
manifolds.
For the  conifolds the complex structures
are apparent in the holomorphic coordinate system that was 
initially  used for the construction of their 
metrics. In Section 3  we shall give the expressions
for their  complex structures 
 in terms of another coordinate system
which is also suitable for deriving the parallel (3,0)-forms.
In Section 4 we shall turn to the case of \CV space 
and derive   
 the corresponding 
 complex structure, K\"ahler form 
    and the parallel (3,0)-form. 
 We will  also give the expression for the associated 
 holomorphic (3,0)-form.
In this way we  will  explicitly 
 verify that the \CV solution preserves
four of the 
supersymmetries of the  type II supergravity  theory, 
in agreement with the arguments presented in \ci{cv,mn}.

It has been shown in  \ci{kt,pt} that
the  solutions representing  configurations
of (ordinary and fractional)  D3-branes 
 on singular \ci{kn,kt}, deformed \ci{kst} and 
resolved \ci{pt} 
conifolds
 can be obtained by solving a system of  first order 
 equations. The latter may be  derived as BPS type of equations
  from a one-dimensional action   admitting a 
 superpotential..
As we shall demonstrate in Section 5, 
the \CV background \ci{cv,mn}
is also a solution of a  a collection of 
first order equations  which
arises as a BPS system   for  a  one-dimensional action
with a superpotential.  This gives another
 indication of a similarity between the 
D3-brane on conifold solutions and the 
 \CV background. In addition, 
this
provides an independent indication   that the
\CV background is supersymmetric.
In the process, we shall  explain  how  all of the solutions  of 
 \ci{kt,kst,pt,mn} can be obtained from a
  single ``interpolating"
 ansatz for the ten-dimensional metric, dilaton 
 and p-form fields.

Appendix A contains some technical  details 
of determining the complex structure of the \CV 
space. In Appendix B 
 supergravity backgrounds  which have 
 the same symmetry as that
 of Calabi spaces are examined. It is found that
 the only such solutions in  the NS$\otimes$NS
 are the Calabi metrics.

\newsection{Supersymmetry 
and  NS$\otimes$NS solutions}

The  type II supergravity field equations that involve
 the  metric $g$,
NS$\otimes$NS  three-form field strength $H$ and the  dilaton $\P$
 are (in the string frame) 
\begin{eqnarray}
& &R_{mn} -{1\over4} H_{mpq}H_n^{\ pq} + 2\nabla_m\partial_n\P =0\ 
,\\
& &\nabla_p\left(e^{-2\P}H^{pmn}\right)=0\ ,\ \ \ \ \ 
\label{st}\end{eqnarray} 
where $\nabla$ is the Levi-Civita connection of the metric $g$ and
$m,n,...=0,\dots,9$. The dilaton equation
$\nabla^2 e^{-2 \P} = { 1 \ov 6} e^{-2 \P} H^{mnk} H_{mnk} $ 
follows from  the above two assuming that the central charge
integration constant vanishes. 

We shall  consider   solutions  of the
type $\bR^{1,3}\times M^6$ which
preserve at least four supersymmetries of the ten-dimensional theory.
We  shall  assume that $H, \P$ 
denote the  restriction of the fields to  $M^6$,
i.e. $H$ is a closed 3-form and $\P$ is a scalar on the
 $M^6$, 
   and the non-trivial part of 
 the ten-dimensional spacetime metric is that of $M^6$. We shall
 use $M,N,K,..=1,2,...,6$ for the indices of $M^6$.
 
 For the background to preserve four supersymmetries, the following
 conditions are required (see \ci{stro,hp}): 

(i) $M^6$ should be  a 
Hermitian manifold with respect to a complex
structure $J$ which is constructed 
from the Killing spinors.

 (ii)  One of the connections with torsion 
$\nabla^\pm=\nabla\pm{1\over2}H$, say $\nabla^+$,
should have its 
holonomy contained in $SU(3)$, i.e. 
\beq
\nabla^+_M J^N{}_P=0\ ,
\ \ \ \ \ \ \ \  R^+_{MN}{}^P{}_Q J^Q{}_P=0\ , 
\label{pteq}
\eeq
where $R^+$ is the curvature of the $\nabla^+$ connection.

 (iii) The K\"ahler form $\Omega$ of $J$ 
 should satisfy 
\beq
{1\over2} \Omega^{MN} H_{MNR}=-2J^K{}_R \partial_K \P\ .
\label{key} 
\eeq
This equation arises
from the dilatino Killing spinor equation. 

In \cite{hp} 
it was shown that
if $J$ is integrable, then the first equation   (\ref{pteq}) is equivalent to
\beq
d\Omega+i_{J} H=0 \ , 
\label{keya}
\eeq
where $i_J$ is the inner derivation with respect to $J$, i.e.
\beq
i_JH_{MNP}=-3 J^K{}_{[M} H_{|K|NP]}\ .
\eeq
Because for the  backgrounds we are considering
the  holonomy of $\nabla^+$ is contained in $SU(3)$,
the manifold $M^6$
admits a   (3,0)-form $\tilde \eta$  which 
is  parallel, i.e.   covariantly constant 
with respect
to the connection $\nabla^+$. Conversely, suppose that $M^6$ is
a hermitian manifold that admits a (3,0)-form. Now if  both the
complex structure and (3,0)-form are parallel with respect
to $\nabla^+$ connection, then the holonomy of $\nabla^+$ is
contained in $SU(3)$ and therefore $M^6$ admits parallel spinors. So
 such background will be supersymmetric provided that some of the
  parallel spinors satisfy
the dilatino Killing spinor equation as well. 
 
Apart from the parallel (3,0)-form $\tilde \eta$,  $M^6$
admits in addition a holomorphic (3,0)-form  $\eta$ given by
\beq
\eta=e^{-2\P} \tilde\eta\ .
\eeq
The existence of $\eta$ 
has been used to show  \cite{ip}  that there are  no  solutions
of this type  with {\it compact} manifolds $M^6$ 
for which $H$ is non-vanishing and the dilaton is a globally
defined scalar on the manifold.
The proof given in \cite{ip} can also be extended, 
after imposing appropriate
boundary conditions, to {\it non-compact} manifolds
$M^6$.  So the
backgrounds for which $\nabla^+$ has holonomy contained in $SU(n)$,
 $H$ is non-vanishing and $\Phi$ is  non-constant
are severely restricted.  Some 
examples in $4k$ dimensions which are not products have been
given in \cite{ptesc}.

It was recently 
 argued  in \ci{mn} that the \CV solution \ci{cv}
provides  another  example of   
NS$\otimes$NS  background with non-compact $M^6$
which preserves four supersymmetries and so the holonomy of $\nabla^+$ is
$SU(3)$.
We shall establish this explicitly in Section 4.

\medskip

A special case arises when $H$ vanishes.
Then  the dilaton $\P$
must be  constant, the manifold $M^6$ is K\"ahler and the holonomy
of the Levi-Civita connection $\nabla$ 
is contained in $SU(3)$, i.e.  
$M^6$ is a
Calabi-Yau manifold.
 The existence of Calabi-Yau metrics for compact and
non-compact manifolds has been established using 
powerful analytical
methods.
 For compact K\"ahler manifolds, the necessary and sufficient
condition required is the triviality of the canonical bundle. 
For
the {\it non-compact}  case, additional information 
regarding boundary conditions
is necessary.
\medskip

 Nevertheless, only 
 very few   non-trivial (non direct-product) 
 examples of such  metrics are known explicitly,
  like the Calabi metrics 
 on the resolved
$\bC^n/\bZ_n$ singularity \ci{who}  and the metrics on 
(singular, deformed and resolved) 
conifolds in  \cite{can,min} and \ci{pt}.
Before   examining   the \CV background, 
 we shall begin in Section 3 with the conifolds
and construct explicitly 
 the corresponding complex structures 
and holomorphic (3,0)-forms. In this way we shall confirm
that the conifolds is a  class of supersymmetric 
backgrounds satisfying the requirements (i)-(iii) above
with vanishing torsion and constant dilaton.
\medskip

Unlike the \CV solution, 
the conifolds preserve eight supersymmetries in the context of
type II theories. This is because for the conifolds $\nabla=\nabla^+=\nabla^-$
since the torsion vanishes. Thus  the holonomy of both $\nabla^+$ and $\nabla^-$
connections is contained is $SU(3)$. In contrast, 
 for the \CV background although 
the holonomy of $\nabla^+$ is $SU(3)$, 
 the holonomy of $\nabla^-$
is $SO(6)$, and,  as a consequence, 
there are no parallel spinors with respect to
$\nabla^-$. Thus the \CV solution preserves half the number
of supersymmetries compared to  the conifolds. 

\medskip

Before we proceed with the case by case investigation, we conclude with
some general remarks regarding the parameterization of almost complex
structures and (3,0)-forms on six-dimensional manifolds. The generalization
to higher dimensions is straightforward. At every point of the
six-dimensional manifold, the almost complex structures that
are compatible with a given metric are parameterized by the
coset space\footnote{Note that $SO(6)/U(3)=\bC P^3$ and 
its dimension is the same as that 
of the manifold. We thank Stefan Ivanov 
for pointing  this out to us.} $SO(6)/U(3)$. Given a metric, it is 
expected that
the associated K\"ahler form of the almost complex structure
is  locally parameterized by six independent functions.
Now,  given a metric and an almost complex structure on $M^6$, at every point
of the manifold the compatible (3,0) forms are parameterized
by the coset $U(3)/SU(3)$. So it is expected that given a frame
of (1,0)-forms associated with the almost complex structure,
the (3,0)-form is determined up to a phase.

\newpage 
\newsection{Complex structure  of  6-d metrics on conifolds}

\subsection{The singular conifold}

\leftline{\underline{K\"ahler structure}}
\vskip 0.3cm

The singular conifold is  the complex three-dimensional subspace
 in $\bC^4$ defined by the
equation \ci{can} 
\beq
\sum_{A=1}^4 (w^A)^2=0\ ,\la{kkk} 
\eeq
where $\{w^A; A=1,2,3, 4\}$ are the coordinates of $\bC^4$.
Clearly, such a space is smooth everywhere apart
from  singular point $w^A=0$.
Let us review the construction of the K\"ahler metric
on the conifold  following 
\ci{can,min}.
Observe that \rf{kkk}  can be rewritten as
\beq
{\rm det}\ W=0\ ,\ \  \ \ \ \ \ \ \ W=w^i\sigma_i+ w^4 {\bf 1}\  , 
\eeq
where  $\{\sigma_i; i=1,2,3\}$
are the Pauli matrices, 
$
\sigma_i \sigma_j= \delta_{ij}{\bf 1}+i \epsilon_{ijk} \sigma_k .
$
Let us  define a  ``radial" coordinate
\beq
\rho^2={\rm Tr} (  W W^\dagger) \ , 
\eeq
and a function $K=K(\rho^2)$ which will be  identified with
 the K\"ahler potential. 
Then the K\"ahler metric on the
conifold is 
\bea
ds^2&=&\partial_\alpha \partial_{\bar \beta}K\  dz^\alpha d\bar z^{\bar\beta}
\nonumber \\
&=&  K''|{\rm Tr}( W^\dagger dW) |^2 +  K'
{\rm Tr} (dW dW^\dagger) \ ,
\label{metr}
\eea
where $ ( ...)' = { d (...) \ov d \rho^2} $ 
and
$\{z^\alpha; \alpha=1,2,3\}$ are some complex coordinates
on the conifold.
The metric can be expressed in terms of five angular coordinates
and the 
radial coordinate $\rho$
by first parameterizing $W$ as
\beq
W=\ \rho \ L_1 Z^{(0)} L_2^\dagger\ , \la{www}
\eeq
where  the $2\times 2$ matrices 
$(L_1, L_2)\in SU(2)\times SU(2)$
 represent the  five angular coordinates,
and
\beq
Z^{(0)}={1\over2} (\sigma_1+i \sigma_2)\ .
\eeq 
To determine the coordinate  dependence of $L_1, L_2$,  it is
convenient to introduce an additional 
coordinate so that the   six  coordinates 
 can be identified with the Euler angles on $SU(2)\times SU(2)$.
Then one can introduce the
  left 
invariant one-forms $\{e^i, \td e^i \}$ \ 
$ (i=1,2,3)$ on 
 $SU(2)\times SU(2)$ group and write
 \bea
L_1^\dagger dL_1= \ha i e^j \sigma_j\ , \ \ \ \ \ \ \ \ \ \ \ \ 
L_2^\dagger dL_2 = \ha  i \tilde e^j \sigma_j\ ,  \eea
so that the  Maurer-Cartan equations are
\beq
de^i=\ha \epsilon^i{}_{jk} e^j\wedge e^k\ ,\ \ \ \ 
\ \ \ \ \ \  \   d\td e^i=\ha \epsilon^i{}_{jk} \td e^j
\wedge \td e^k\ .\la{minu}
\eeq
The computation of the metric and the K\"ahler form
proceeds by using  the properties of the
left-invariant forms and by treating  the 
two of the six angular coordinates,  which correspond to
the ``third" Euler angles, as independent,  
though the final result will effectively
depend only on their sum $\psi$
  (see \ci{min}).
As a result, one finds the metric 
\bea
ds^2&= &(\rho^2 K''+K') d\rho^2+\fo (\rho^4 K''+\rho^2 K') (e^3+\tilde e^3)^2
\nonumber \\
&+& \  \fo 
K' \rho^2 \big[ (e^1)^2+(e^2)^2+(\tilde e^1)^2
+(\tilde e^2)^2\big]\ .
\label{kaehler}
\eea
The frames that appear in the expression for the metric are
\bea 
e^1&=&  \sin{\psi\ov 2} \sin\theta_1 d\phi_1+\cos{\psi\ov 2}
d\theta_1\ , \nonumber \\
e^2& =& - \cos{\psi\ov 2}\sin\theta_1 d\phi_1+\sin{\psi\ov 2} d\theta_1
\ ,  \nonumber \\
\tilde  e^1&=&  \sin{\psi\ov 2} \sin\theta_2d\phi_2+\cos{\psi\ov 2}
d\theta_2\ , \nonumber \\
\tilde e^2 &=& - \cos{\psi\ov 2}\sin\theta_2 d\phi_2+\sin{\psi\ov 2} d\theta_2
\ ,  \nonumber \\
e^3 + \tilde e^3 &= &d\psi 
+ \cos\theta_1 d\phi_1 + \cos\theta_2 d\phi_2  \ .
  \la{basi}
     \eea
The K\"ahler form of the conifold is defined as 
\bea 
\Omega={i\over2} \partial\bar \partial K =  i \bigg[{\rm Tr} 
(W^\dagger dW) \wedge\rho d\rho K''+{1\over2}{\rm Tr} (dW\wedge 
dW^\dagger) K'\bigg]\ . \label{kaehlerf} \eea 
Using the  above 
relations, 
  $\Omega$ can be expressed in terms of the coordinates as
\beq
\Omega={1\over4} \bigg[ d(\rho^2 K')\wedge (e^3+\tilde e^3)
+ \rho^2 K' (e^1\wedge e^2+
\tilde e^1\wedge \tilde e^2)\bigg] . \la{kah}
\eeq
 From the expressions \rf{kaehler},\rf{kah} 
it is straightforward
 to find
the adapted frame for this K\"ahler structure,
i.e.  the frame in which both the metric and  $\Omega$ 
take the standard forms with {\it constant}  coefficients
\bea
\ae^1&=& \ha (\rho^2K')^{{1\over2}} e^1 , 
 \ \ \  \ae^2=\ha (\rho^2 K')^{{1\over2}} e^2 , \ \ \ 
\ \ \ 
\ae^3=\ha (\rho^2 K')^{{1\over2}}\tilde e^1,  \ \ \ 
\ae^4= \ha (\rho^2 K')^{{1\over2}}
\tilde e^2 , \nonumber \\
\ae^5&=& (\rho^4 K''+\rho^2 K')^{{1\over2}} \rho^{-1}  d\rho\ , 
\qquad  \ae^6=\ha 
(\rho^4 K''+\rho^2 K')^{{1\over2}} (e^3+\tilde e^3)
\ . \la{fra}  \eea
In particular,  the 
complex structure in this frame is
\bea
J(\ae^1)&=&-\ae^2, \qquad J(\ae^3)=-\ae^4,\qquad J(\ae^5)=-\ae^6, 
\nonumber\\
J(\ae^2)&=&\ae^1, \qquad \ \ J(\ae^4)=\ae^3,\qquad \ \ \ J(\ae^6)=\ae^5\ . 
\la{stra} 
\eea 
It can be verified by a straightforward computation
that the above complex structure is  integrable as it
is expected. The most convenient way to show this 
is by using the Frobenius theorem; see Section 4 for details.

\newpage
\leftline{\underline{Calabi-Yau structure}}
\vskip 0.3cm

The above class of K\"ahler metrics on the
conifold contains a Calabi-Yau 
one, i.e. a metric  that it is Ricci flat and so it has holonomy $SU(3)$. One way to
find it  is by constructing the  parallel
(3,0)-form $\eta$ associated with  such manifolds as has been explained
in Section 2.
Indeed,  we take
\beq
\eta= (\ae^1+i\ae^2)\wedge (\ae^3+i\ae^4)\wedge (\ae^5+i\ae^6)\ .
 \la{foo}
\eeq
It turns out that there is no need to add a phase in the (1,0) frame
$\{\ae^1+i\ae^2, \ae^3+i\ae^4, \ae^5+i\ae^6\}$ we have chosen.\footnote
{However a phase would have been necessary 
if another (1,0) frame was chosen instead.}
This is clearly a (3,0)-form. It turns out that it is sufficient
to check that it is closed as well. This is the case   
provided 
\beq
\rho^2 G'-G=0\ , \la{ff}
\eeq
where
\beq 
G\equiv \rho^2 K' (\rho^4 K''  +   \rho^2 K')^{1\over2}\ . \la{fgfg}
\eeq
Eq. \rf{ff}  can be integrated to give
\beq
{G\over \rho^2}=\lambda=\const\ , \la{fff} 
\eeq
where $\lambda$ is a  positive constant;
for $\l=0$ the metric is degenerate. 
Integrating \rf{fff}, one finds
\beq
K'=\big( {3\lambda^2\over 2\rho^2}+{c\over \rho^6}\big)^{1\over3}
\ , \label{kk}
\eeq
where $c$ is another integration constant.
There is no need to
integrate once more to determine  $K$ since 
the metric, the complex structure and
the (3,0)-form  all depend on the   derivatives of the K\"ahler
potential. Only one combination of $\l$ and $c$ is a nontrivial
parameter,  so the K\"ahler potential  (\ref{kk}) thus 
determines
a {\it one parameter}  family of Calabi-Yau  metrics
on the conifold.
 
For $c=0$, the K\"ahler potential is
\beq
K=\big({\textstyle {9\lambda\over4}}\big)^{2\over3} \rho^{4\over3}\ . 
\eeq
 After a redefinition 
of the $\rho$ coordinate (or setting $\l^2= 2/3$), \rf{kaehler} 
 becomes the  standard Calabi-Yau metric on the  cone
over the homogeneous space $T^{(1,1)}$, i.e.  a coset $\big[SU(2)\times
 SU(2)\big]/U(1)$ \ci{pop}. This 6-d metric
is singular at  the apex   $\rho=0$ of the cone.
It is natural to restrict the
 parameter $c$  to positive values
so that the metric remains regular for  all values
of $\rho^2$ apart from 
 $\rho^2\rightarrow 0$.
 
\newpage
\subsection{The resolved conifold}

\leftline{\underline{K\"ahler structure}}
\vskip 0.3cm

Singularities
like that at the apex of the singular conifold in the previous
section can be removed  either by blowing up the singular point or
by deforming the associated set of algebraic equations.
The former case will be investigated in this
subsection while  the latter  will be examined in the next one.

The discussion of the complex structure 
of 6-d metrics on resolved conifold \ci{can}
 is parallel to the one in the singular conifold case above.
The K\"ahler potential is\foot{Here we use $\r$
instead of $r$ in \ci{pt} for the radial coordinate.}
 \beq
  K = F (\r^2)+4 a^2 {\rm ln} (1+|\Lambda|^2)\ , 
 \eeq
 where
 \beq
 \r^2=(1+|\Lambda|^2) (|U|^2+|Y|^2)\ .
 \eeq
 Here $(U,Y,\Lambda)$ are the holomorphic coordinates which can be
 parameterized
 in terms of the Euler angles as
 $$
 U= \r e^{{i\over2} (\psi+\phi_1+\phi_2)}
  \cos {\theta_1\over2} \cos{\theta_2\over2}
 \ , \ \  
 Y= \r e^{{i\over2} (\psi-\phi_1+\phi_2)}
  \sin {\theta_1\over2} \cos{\theta_2\over2}
  \ , \ \ 
  \Lambda= e^{-i\phi_2}\tan{\theta_2\over2}\ .  $$
The metric of the resolved conifold  can
  then  be written as \ci{pt}  (cf. \rf{kaehler})
\bea
ds^2&=&(\rho^2 F''+F') d\rho^2+\fo
(\rho^4 F''+\rho^2 F') (e^3+\tilde
e^3)^2
\nonumber \\
&+& \ 
\fo F' \rho^2 \big[ (e^1)^2+(e^2)^2]
+\fo  ( F' \rho^2  + 4a^2) [(\tilde e^1)^2+(\tilde e^2)^2\big]\
, 
\label{hler}
\eea
where the 1-forms are expressed in terms of the Euler angles as 
in \rf{basi}.
This  metric has explicit $SU(2)\times SU(2) $
invariance  and 
 becomes the same as in \rf{kaehler}  when  $a=0$
 (and thus $K=F$).\foot{For small $\r$
 the
$S^3$ ($\psi,\theta_1,\phi_1$)  part of the metric
 shrinks to zero size while the $S^2$ $(\te_2,\p_2)$ part
stays finite with  radius $a$.}
The   dependence on the resolution parameter 
$a$  here is  explicit,
 but after imposing additional conditions,
like 
Ricci-flatness, $F$ may  start depending   on $a$.

The K\"ahler form of the resolved conifold is then (cf. \rf{kah}) 
\bea
\Omega={i\over2} \partial\bar \partial  K
={1\over4} \bigg[ d(\rho^2 F')\wedge (e^3+\tilde e^3)+ \rho^2
F'  e^1\wedge e^2+ ( \rho^2 F' + 4 a^2) 
\tilde e^1\wedge \tilde e^2\bigg] . \la{kahe}
\eea
The adapted frame for the metric and the K\"ahler form   
 is (cf. \rf{fra}) 
\bea
\ae^1&=& \ha (\rho^2F')^{{1\over2}} e^1 \ , \ \ \ \ \ \ \ \ \ 
 \ \ \ \ \ \ \ \ \  \ae^2=\ha (\rho^2 F')^{{1\over2}} e^2 \ ,\nonumber \\
\ae^3&=&\ha (\rho^2 F' + 4 a^2 )^{{1\over2}}\tilde e^1\ , \ \ \ \ 
\ \ \  \ \ \ 
\ae^4= \ha (\rho^2 F' + 4 a^2 )^{{1\over2}}
\tilde e^2 \  , \nonumber \\
\ae^5&=& (\rho^4 F''+\rho^2 F')^{{1\over2}} \rho^{-1}  d\rho\ , 
\ \ \  \ \ \   \ae^6=\ha 
(\rho^4 F''+\rho^2 F')^{{1\over2}} (e^3+\tilde e^3)
\ . \la{riuy} \eea
The 
complex structure in this frame  takes the same form 
as in \rf{stra} and is integrable.


\vskip 0.5cm
\leftline{\underline{Calabi-Yau structure}}
\vskip 0.3cm

To find the Calabi-Yau representative in the class of metrics 
\rf{hler} we again construct the  holomorphic
(3,0)-form $\eta$. The required  form has the same structure 
 as in \rf{foo} 
\beq
\eta= (\ae^1+i\ae^2)\wedge (\ae^3+i\ae^4)\wedge (\ae^5+i\ae^6)\ .
 \la{fooe}
\eeq
It is closed, 
provided 
\beq
\rho^2 G'-G=0\ , \la{feef}
\eeq
where (cf. \rf{fgfg}) 
\beq 
G\equiv 
[ \rho^2 F' (\r^2 F' + 4 a^2)  (\rho^4 F'' + \rho^2 F')]^{1\over2}\ .
\eeq
Thus, as in  \rf{fff},  
$
{G\over \rho^2}=\lambda=\const.$
Integrating once more, we get 
\bea
(\rho^2 F')^3 +  6 a^2 (\rho^2 F')^2  = 
 { 3 \ov 2} \l^2  \r^4  + c  \ , 
\eea
which  reduces to \rf{kk} for $a=0$.
This is the same as the equation on $F$
 found in \ci{pt} from the
Ricci-flatness condition (in \ci{pt} the constants were chosen 
to be  $\l^2  = 2/3$ and $c=0$). 
Thus, in general,  we get  
a one (non-trivial) 
 parameter family of Calabi-Yau  metrics
on the resolved conifold.

\subsection{The deformed conifold}
\leftline{\underline{K\"ahler structure}}
\vskip 0.3cm
\def \a {\alpha}
\def \b {\beta}

In the deformed conifold case \ci{can}
the singular point $w^A=0$ is excluded
by  replacing \rf{kkk} by 
\beq
{\rm det}\ W=-{\epsilon^2\over2}\ ,
\eeq
where $\epsilon$ is a  ``deformation" parameter.
 Again, one  parameterizes $W$ as \ci{min} 
\beq
W=\ \rho \ L_1 Z^{(0)}_\epsilon L_2^\dagger\ ,
\label{nw}
\eeq
where now (cf. \rf{www}) 
\beq
Z^{(0)}_\epsilon=\ha \a (\rho) ( {\sigma_1+i\sigma_2}) +
\ha \b (\rho)  ({\sigma_1-i\sigma_2}) \ , 
\eeq
\bea
\a\equiv  {1\over2} \big(\sqrt{ 1+{\epsilon^2\over\rho^2}}
+\sqrt {1-{\epsilon^2\over\rho^2}}\big)\ , \ \ \ \ \ 
\b\equiv  \sqrt {1 - \a^2}  = {\epsilon^2\over 2\rho^2} \a^{-1}\ .
\la{aabb}
\eea
Observe that $\a^2+\b^2=1$.
The metric and K\"ahler form are determined by $W$ 
through  the general expressions 
 (\ref{metr})
and (\ref{kaehlerf}).
To compute them, we again  organize the
angular variables
in terms of the two sets of $SU(2)$ Euler angles and
  left-invariant forms $\{e^i, \tilde e^i\}$ as in
   the singular conifold case.
Then the metric of the deformed conifold
\bea
ds^2=&& K''\rho^2 d\rho^2 +K' \big(d\rho^2+\rho^2d\a^2+\rho^2 
d\b^2\big)
\nonumber\\
&+& \fo \big[\rho^4 (\a^2-\b^2) K''+\rho^2 K' \big] (e^3+\tilde e^3)^2
\la{mee} \\
&+& \fo K' \rho^2 \bigg[ (\a e^1-\b\tilde e^1)^2+(\b e^1-\a 
\tilde e^1)^2
+ (\a e^2+\b\tilde e^2)^2+(\b e^2+\a \tilde e^2)^2\bigg]\ ,
\nonumber\eea
or,  equivalently (cf.  \rf{kaehler},\rf{hler}) 
\bea
ds^2= \big[K''(1-{\epsilon^4\over \rho^4})\rho^4+ K' \rho^2\big]
 \bigg[{d\rho^2\over \rho^2(1-{\epsilon^4\over \rho^4})}
 +\fo (e^3+\tilde e^3)^2\bigg]
\nonumber \\
+ \ \fo   K' \rho^2 \bigg[ ( \a e^1-\b\tilde e^1)^2+(\b e^1-
\a \tilde e^1)^2
+ (\a e^2+\b\tilde e^2)^2+(\b e^2+\a \tilde e^2)^2\bigg]\ .
\la{demm}
\eea
Similarly, the K\"ahler form of the deformed conifold is 
\beq
\Omega={1\over4}\bigg[ d\big[\rho^2K' (\a^2-\b^2)\big]
\wedge (e^3+\tilde e^3)+ \rho^2 K' (\a^2-\b^2)
(e^1\wedge e^2+\tilde e^1\wedge \tilde e^2)\bigg]\ .
\eeq 
These expressions reduce to the ones in the singular conifold case
in the  $\ep=0$ limit. 

The adapted frame of this K\"ahler structure
where the metric and the  K\"ahler form have  constant
components
is the following (cf. \rf{fra},\rf{riuy} ):
\bea
\ae^1&=&\ha (K' \rho^2)^{{1\over2}} (\a e^1-\b\tilde e^1)\ ,\qquad 
\ae^2=\ha (K' \rho^2)^{{1\over2}} (\a e^2+\b \tilde e^2)\ , 
\nonumber\\
\ae^3&=&\ha (K' \rho^2)^{{1\over2}} (-\b e^1+\a\tilde e^1)\ ,\qquad
\ae^4=\ha (K' \rho^2)^{{1\over2}} (\b e^2+\a \tilde e^2)\ , 
\nonumber\\
\ae^5&=&\big[K''(1-{\epsilon^4\over \rho^4})\rho^4+
 K' \rho^2\big]^{1\over2}
{d\rho\over \rho(1-{\epsilon^4\over \rho^4})^{1\over2}}\ , 
\nonumber\\
\ae^6&=&\ha \big[K''(1-{\epsilon^4\over \rho^4})\rho^4+ 
K' \rho^2\big]^{1\over2}
(e_3+\tilde e_3) 
\la{forr}\ .
\eea
Indeed, like in the singular and resolved conifold cases, 
in this adapted frame  the K\"ahler form becomes simply 
\beq
\Omega=\ae^1\wedge \ae^2+\ae^3\wedge \ae^4+\ae^5\wedge \ae^6\ .
\eeq
The complex structure is thus 
\bea
J(\ae^1)&=&-\ae^2\ ,  \qquad J(\ae^3)=-\ae^4\ 
,\qquad J(\ae^5)=-\ae^6\ , 
\nonumber\\
J(\ae^2)&=&\ae^1\ , \qquad \  \ J(\ae^4)=\ae^3 \ ,\qquad \  \ \  
J(\ae^6)=\ae^5\ .
\eea 
\vskip 0.3cm

\leftline{\underline{Calabi-Yau structure}}
 \vskip 0.3cm 
 
The class of  K\"ahler metrics  on the deformed conifold
includes also  
 a Calabi-Yau metric. As  in the case of the singular
  conifold \rf{foo},
 one can define the 
(3,0)-form by 
\beq 
\eta=(\ae^1+i\ae^2)\wedge (\ae^3+i\ae^4)\wedge (\ae^5+i\ae^6)\ .
 \eeq 
 For  the K\"ahler structure
to be associated with the Calabi-Yau metric, the form
$\eta$  must be  closed. This is
the case provided that (cf. \rf{ff}--\rf{fff}) 
\beq
{\a \b^2\over \a' (\a^2-\b^2)} G'- G =0\ ,
\ \ \ \  {\rm i.e.} \ \ \ \    \r^2 G'- G=0 \ ,  \la{fef} 
 \eeq
 \beq
 G\equiv \rho^2 K'
 \bigg[  (1-{\epsilon^4\over \rho^4}) \rho^4 K'' +   \rho^2 K' 
  \bigg]^{1\over2} 
 \ .
 \eeq
 We have used the expressions \rf{aabb} 
  for $\a,\b$ \rf{fef} to show that the equation 
  for $G$ is the same as in  \rf{ff},\rf{feef}. 
  Integrating it, 
 we find  again 
 \beq 
 {G\over \rho^2}= \l=\const\ , 
 \eeq
 where $\l$  is assumed to be  positive. 
 Integrating this equation once more, 
 we get 
 \beq
 K'={1\over \sqrt{\rho^4-\epsilon^4}} 
 \bigg[ { 3\ov 2}  \l^2 \r^2 \sqrt{\r^4 - \epsilon^4}  +  c 
     - { 3\ov 2}  \l^2 \epsilon^4
     \log( \r^2 + \sqrt{\r^4 - \epsilon^4} ) \bigg]^{1 \ov 3}  \ , 
 \label{kkd}
 \eeq
 where $c$ is another integration constant
 (again, only one combination of $c$ and $\l$ is a non-trivial
 parameter).
 This expression reduces to \rf{kk} in the $\ep\to 0$ limit.
 As in the conifold case, finding  the derivative $K'$ 
  of the
 K\"ahler potential with respect to $\rho^2$
  suffices to determine the metric, complex structure
 and the holomorphic (3,0)-form. The K\"ahler potential
 following from  (\ref{kkd}) 
  defines a one-parameter family of Calabi-Yau
 metrics on the deformed conifold.

\def \te {\theta}

\newsection{ The Complex Geometry  of  \CV  background}
\subsection{The solution}

The NS$\otimes$NS solution of \ci{cv} lifted to ten dimensions
has 
 form $\bR^{1,3} \times M^6$, where
  $M^6$ has  topology
 $\bR \times S^2 \times S^3$. The string frame metric
 is the direct sum of the $\bR^{1,3}$ and $ M^6$ metrics.
The metric on $M^6$, the 3-form 
 and  the dilaton  are given by \ci{mn} 
\bea
ds^2&=& dr^2+ e^{2g(r) } (d\theta^2+\sin^2\theta
 d\varphi^2)+{1\over4}
 \sum_{i=1}^3(\epsilon^i-A^i)^2
 \ , \la{mert}\\
 H&=&-{1\over4} (\epsilon^1-A^1)
 \wedge(\epsilon^2-A^2)\wedge(\epsilon^3-A^3)
 +{1\over4}
 \sum_{i=1}^3 F^i \wedge (\epsilon^i-A^i) \ , 
 \\
 e^{2\P}&=&e^{2\P_0}{2e^{g } \over \sinh 2r }\ , \la{dila}
 \eea
 respectively,
 where
 \bea
 A^1= a(r) d\theta\ , \ \ \ \ \ 
 A^2=  a(r) \sin\theta d\varphi
 \ ,  \ \ \ \ \ \ 
 A^3=  \cos\theta d\varphi\ , 
 \eea
 and
 \bea
 a= {2r\over \sinh 2r }\ , \ \ \ \ \ \ \ \ \  
 e^{2g}=  r \coth 2r -{r^2\over \sinh^2 2r }-{1\over4}\ .\la{gaa}
 \eea
$F$ is the  curvature of the connection $A$,  i.e. 
$F^i=dA^i +{1\over2} \epsilon^i{}_{jk} A^j \wedge A^k$, 
and $\epsilon^i $ are the left-invariant one-forms 
on $S^3$ which now satisfy (cf. \rf{basi},\rf{minu})  
\beq
d\epsilon^i=-{1\over2} \epsilon^i{}_{jk}
 \epsilon^j\wedge \epsilon^k\ . \la{noww}
 \eeq
It is convenient to define the frame 
in which the metric is diagonal 
\bea
\xi^0=dr \ ,\ \ \ \xi^1=e^g d\theta\ , \ \ \  \xi^2=e^g \sin\theta d\varphi\ ,
\ \ \ \ 
\rho^i={1\over2} (\epsilon^i-A^i )\ .
\eea
The 3-form $H$ then  becomes
\beq
H=-2 \rho^1\wedge \rho^2\wedge \rho^3+{1\over2} \sum_{i=1}^3 
F^i \rho^i\ .
\eeq
In addition, we have
\bea
A^1= a e^{-g} \xi^1,\qquad 
A^2= a  e^{-g} \xi^2,\qquad 
A^3=   e^{-g}\cot\theta \xi^2\ , 
\eea
and $
F^{1,2}=  e^{-g} da\wedge \xi^{1,2} , \ \
F^3=  (a^2-1) e^{-2g} \xi^1\wedge \xi^2 .$
Observe also that
\bea
d\xi^1= dg\wedge \xi^1
\ , \ \ \ \ \ \ 
d\xi^2=dg\wedge \xi^2+ e^{-g} \cot\theta\, \xi^1\wedge \xi^2 \ , 
\eea
\beq
d\rho^i=-\epsilon^i{}_{jk} \rho^j\wedge \rho^k-
 \epsilon^i{}_{jk}A^j \rho^k-{1\over2} F^i\ .
\eeq

\subsection{Conditions on  the  K\"ahler form}


To find the complex structure associated with the the
\CV manifold, we consider the  most general candidate
for a  K\"ahler two-form 
\beq
\Omega= \lambda_i \xi^0\wedge \rho^i+  w_a dr\wedge \xi^a+z_{ia} \rho^i\wedge e^a
+{1\over2} \mu^i\epsilon_{ijk} \rho^j\wedge \rho^k+{1\over2} 
P\epsilon_{ab} \xi^a\wedge \xi^b\ , \la{ansa}
\eeq
where  $i,j=1,2,3$ and $ a,b=1,2$
  are frame indices raised
and lowered with  flat Euclidean metric.
The components  $(\lambda_i, w_a,  z_{ia}, \mu^i, P$) of the form
$\Omega$ are allowed to depend on all of  the
coordinates of the manifold $M^6$.

First,  let us find the conditions on $\Omega$ 
such that
it can be identified with 
the K\"ahler form of an almost complex structure $J$. 
For this
$J^A{}_B=\delta^{AC} \Omega_{CB}$ should satisfy $J^2=-1$
($A,B,C=1, \dots, 6$
are frame indices).
It is best to perform the calculation by taking (cf. \rf{ansa}) 
\bea
i_J(\xi^0)&=&-\lambda_i \rho^i-w_a \xi^a\ , 
\nonumber\\
i_J(\rho^i)&=& \lambda^i \xi^0-z^i{}_a \xi^a-\mu^j \epsilon_j{}^i{}_k 
\rho^k\ , 
\nonumber\\
i_J(\xi^a)&=& w^a \xi^0+z_i{}^a \rho^i-P \epsilon^a{}_b \xi^b\ , 
\eea
and then verifying that $i_J i_J=-1$.
Acting on $dr$ twice with $i_J$, we get
the following conditions 
\bea
\lambda_i \lambda^i+w_a w^a&=&1\ , 
\nonumber\\
\lambda_i z^i{}_a+Pw_b \epsilon^b{}_a&=&0\ , 
\nonumber\\
-\lambda_i \mu^j \epsilon_j{}^i{}_k+ z_k{}^a w_a &=&0\ . 
\label{oo}
\eea
Similarly, from  $i_J( i_J(\rho^i))=-\rho^i$, we get
\bea
\lambda_i \lambda_j+z_{ia} z_j{}^a+
 \mu^k \mu_k \delta_{ij}-\mu_i \mu_j&=&\delta_{ij}\ , 
 \nonumber\\
 -\lambda_i w_a+P z_{ib } \epsilon^b{}_a
 +\mu^k \epsilon_k{}^{ij} z_{ja}&=&0\ .  
 \label{ooo}
 \eea
 Finally,  acting on $\xi^a$, we find 
 \beq
 w_a w_b+z_{ia} z^i{}_b+ P^2 \delta_{ab}=\delta_{ab}\ . 
 \label{oooo}
 \eeq
 General solution of the algebraic constraints 
 \rf{oo},\rf{ooo},\rf{oooo} is given in Appendix A.
 
 It remains  to find the differential equations
 on the components of $\Omega$ \rf{ansa} 
 such that the non-trivial 
  conditions in (\ref{key}) and (\ref{keya}) are satisfied.
 Eq. (\ref{key})  leads to 
 \bea
 z_{11}+z_{22}=0\ , 
 \nonumber \\
 -4\mu_1+ \partial_r a e^{-g} w_1=-4 \lambda_1 \partial_r\P\ , 
 \nonumber\\
  -4\mu_2+ \partial_r a e^{-g} w_2=-4 \lambda_2 \partial_r\P\ , 
  \nonumber\\
   -4\mu_3+(a^2-1) e^{-2g} P=-4 \lambda_3 \partial_r\P\ , 
 \nonumber\\
 (a^2-1) e^{-2g} z_{32}+\partial_r a e^{-g} \lambda_1=+4 w_1
  \partial_r\P\ , 
 \nonumber\\
  (a^2-1) e^{-2g} z_{31}-\partial_r a e^{-g} \lambda_2=-
  4 w_2 \partial_r\P\ , 
  \label{iii}
\eea
where $z_{11}, z_{22}, z_{31}$ and $z_{32}$
are the components of $z_{ia}$ in \rf{ansa}.
 Evaluating $d\Omega$ and $i_JH$ we find 
from  (\ref{keya}) the following additional conditions
\bea
-D_a \lambda_i+D_i w_a
+\lambda_k \epsilon^k{}_{ji} A^j{}_a- e^{-g} \partial_r(z_{ia} e^g)
- \mu^k \epsilon_{kji} F^j_{ra}~~~~& &
\nonumber\\
-\ {1\over2} F^i_{rb} P \epsilon^b{}_a+{1\over2}
F^i_{ba} w^b&=&0\ , 
\nonumber \\
-\  (D_j\lambda_k-(k,j))
+ \partial_r \mu_i \epsilon^i{}_{jk}
+{1\over 2 }\big( z_j{}^a F^k{}_{ra}-(k,j)\big)&=&0\ , 
\nonumber \\
 \lambda_3 F^3_{ab}
- (D_a w_b-D_bw_a)-{1\over2} w_2
\cot\theta e^{-g} \epsilon_{ab}~~~~~ & &
\nonumber\\
+\  \partial_rP \epsilon_{ab}+P\partial_rg
\epsilon_{ab}- \big(F^i_{ra} z_{ib}-(b,a)\big)&=&0\ , 
\nonumber\\
 D_iP\epsilon_{ab}+\big(D_a z_{ib}
-z_{kb} \epsilon^k{}_{ji} A^j{}_a-(b,a)\big)
+ z_{i2}
\cot\theta e^{-g} \epsilon_{ab}~~~~~& &
\nonumber\\
+\  \mu^k \epsilon_{k3i} F^3_{ab}-
{1\over2} \big(F^i_{ra} w_b-(b,a)\big)&=&0\ , \nonumber
\\
(D_j z_{ka}-(k,j))+ D_a \mu_i \epsilon_{ijk}
\nonumber\\
+\  \big(-\mu_j A^k_a 
+{1\over2} F^k_{ra} \lambda_j
-{1\over2} F^k_{ba} z_j{}^b -(j,k)\big)=0\  , 
\label{td}
\eea
where the derivatives $D_a$ and $D_i$ are defined by the
equation $df=dr \partial_rf+e^a D_af+\rho^i D_if$
for any  function $f$ on $M^6$.

We shall now solve the equations (\ref{oo})--(\ref{oooo}), (\ref{iii})
and (\ref{td}) by assuming that the components of the K\"ahler form
$\Omega$ depend only on the {\it radial coordinate} $r$. 
Using this assumption,  from (\ref{td}) one finds that
\bea
\lambda_1=\lambda_2=w_2=z_{32}=z_{11}=z_{22}=\mu_1=\mu_2&=&0\ , 
\nonumber\\
z_{21}+z_{12}&=&0\ . \la{aa}
\eea
{}From the last two equations in (\ref{iii}), we find also that
\beq
w_2=z_{31}=0\ . \la{aaa}
\eeq 
The condition (\ref{oo}) implies that $\lambda_3=\pm 1$. We shall
choose in what follows $\lambda_3=1$.
Setting $z_{12}=-z_{21}\equiv X$, the equations
 (\ref{oo})--(\ref{oooo}) imply that
\beq
\mu_3=-P\ , \ \ \ \qquad X^2+P^2=1\ .
\label{squa}
\eeq
The remaining  equations in (\ref{iii}) and (\ref{td}) give an
 over-determined
linear system on  $P$ and $X$. In particular,
 $P$ can be found 
in terms of $a, g$ and $\P$ from the third equation in (\ref{iii}).
Simplifying the rest of the equations using $X X'+PP'=0$ 
(which follows from 
(\ref{squa})), the final set of equations is the following:
\bea
P= -{4\P'\over 4+(a^2-1) e^{-2g}} &,&
\\
X={P'\over a' e^{-g}}&,&
\\
X'=-a' e^{-g}P&,&
\\
{1\over2} (a^2-1) e^{-2g}-{1\over2} a' e^{-g} X+P g'=0&,&
\\
a e^{-g} +{1\over2} Pa' e^{-g}+X g'=0&,&
\\
-Pa-{1\over2} a'+{1\over2} (a^2-1) e^{-g} X=0&,&
\label{fset}
\eea
where $X'=\partial_rX$ and  $P'=\partial_rP$.
The first two equations determine the remaining
two unknown functions
$P$ and $X$. The rest of the equations are satisfied 
automatically 
by using  the expressions for the functions 
$a,g,\P$  given in \rf{dila},\rf{gaa}.   

To summarize, the {\it K\"ahler
form}  for the  solution of \ci{cv,mn} is
\beq
\Omega=\xi^0 \wedge \rho^3+X(r) (\rho^1 \wedge \xi^2-\rho^2\wedge \xi^1) 
 +P(r)  (- \rho^1\wedge \rho^2+\xi^1\wedge \xi^2)\ , \la{kak} 
\eeq
where 
 \bea
P= {  \sinh 4 r  - 4 r\ov  2 \sinh^2 2 r } \ , \ \ \ \ \ \ \ \ \ 
X = ( 1 - P^2)^{1/2}  \ . \la{resor}
\eea
It remains to show that the almost complex structure associated with
this  K\"ahler form  \rf{kak}  is {\it integrable}. 
This can be  verified  by introducing  the
(1,0)-forms
\bea
\Epsilon^1&=&\xi^0+i\rho^3\ , 
\nonumber\\
\Epsilon^2&=&\rho^1+i X \xi^2-i P \rho^2\ , 
\nonumber\\
\Epsilon^3&=& \xi^1+i X \rho^2+i P \xi^2\ .
\eea
Then according to the Frobenius theorem, the almost complex
structure $J$ is integrable provided that the (0,2) part
of the 2-forms $d\Epsilon^1, d\Epsilon^2$ and $d\Epsilon^3$
vanishes. This requirement leads to the following conditions:
\bea
X+a e^{-g} P-{ 1\over 4} (a^2-1) e^{-2g} X&=&0\ , 
\nonumber \\
a e^{-g} X+ X^2 g'+{1\over2} e^{-g} PX a' &=&0\ , 
\nonumber\\
-X-a e^{-g} P-{1\over4} e^{-g} a'- {1\over2} PX'~~~~~~~& &\ , 
\nonumber \\
+{1\over2} XP'-
{1\over2}XPg'-{1\over4} e^{-g} P^2 a'&=&0\ , 
\nonumber\\
{1\over2} g'+{1\over2}a e^{-g} X+{1\over4} e^{-g} PX a'-{
1\over2} P^2 g'&=&0\ , 
\nonumber\\
X+{1\over2}PX'-{1\over2}XP'+{1\over4} e^{-g} X^2 a'-
{1\over2} PX g'&=&0\ ,
\eea
 which are satisfied automatically. 
 Since 
 $d\Omega+i_JH=0$ is
satisfied and the almost complex structure is integrable, then
as it has been explained in Section 2, $J$ is parallel with
respect to the $\nabla^+$ connection. Thus the first condition
in (\ref{pteq}) is satisfied.

It remains to prove that the second condition in (\ref{pteq})
 is satisfied as well. For this,
  it can be shown using  various identities in \ci{ip}
 that
if a background satisfies: 
 (i) the field equations, 
(ii) admits a complex structure
which is parallel with respect to the $\nabla^+$ connection and
(iii) obeys
the Killing spinor equation associated with the dilatino,
then the holonomy of the $\nabla^+$ connection is contained in $SU(n)$.
The \CV background has been shown to obey all these
three requirements and so it also satisfies the second condition (\ref{pteq}).
{}It follows from all the above  that the \CV background
preserves at least four supersymmetries.
It can be easily verified that the $\nabla^-$ connection does not
admit a parallel complex structure with similar properties.
Therefore, 
 the holonomy of $\nabla^-$ is $SO(6)$. We
conclude that the \CV background preserves four supersymmetries
 (in agreement
with the arguments in \ci{cv,mn}).

The $\nabla^+$-parallel (3,0) form for the \CV background
is
\beq
\tilde \eta = \Epsilon^1\wedge \Epsilon^2\wedge \Epsilon^3\ , 
\eeq
and the associated holomorphic (3,0) form is
\beq
\eta = e^{-2\P}\Epsilon^1\wedge \Epsilon^2\wedge \Epsilon^3\ .
\eeq
Let us note that in the asymptotic limit 
 $r\rightarrow \infty$, the \CV  solution simplifies 
 considerably, with the functions in \rf{dila}
 becoming  
\bea
\P= \P_0 -r+{1\over4} \ln r \ , \ \ \ \  \  a=0\  , \ \ 
\ \ \ 
e^{2g} = r\  , \la{asa}  
\eea
up to exponentially suppressed corrections.
In this limit   $P=1$ and $X=0$ (see \rf{resor})
 so that the K\"ahler
form     \rf{kak}        reduces to 
\bea
\Omega&=& \xi^0\wedge \rho^3-\rho^1\wedge
 \rho^2+ \xi^1\wedge \xi^2
\nonumber\\
&=&{1\over2} dr\wedge (\epsilon^3-\cos\theta d\varphi)-
{1\over4}\epsilon^1\wedge \epsilon^2+ r \sin\theta
d\theta\wedge d\varphi\ . \la{assa} 
\eea
Finally, the $\nabla^+$-parallel (3,0)-form is
\beq
\tilde \eta= {{\sqrt r}\over2} \big[dr+ {i\over2}
 (\epsilon^3-\cos\theta d\varphi)\big]
\wedge (\epsilon^1-i\epsilon^2)\wedge
 (d\theta+i\sin\theta d\varphi)\ , 
\eeq
while the holomorphic (3,0) form is
\beq
\eta= {1\over2} e^{2r}  \big[dr+ {i\over2} 
(\epsilon^3-\cos\theta d\varphi)\big]
\wedge (\epsilon^1-i\epsilon^2)\wedge (d\theta+i\sin\theta d\varphi)\ .
\eeq
\vskip 1cm


\def \ov {\over}
\def \ha { { 1\ov 2}}
\def \we { \wedge}
\def \P { \Phi} \def\ep {\epsilon}

\def \tv   {{1 \ov 12}}
\def \go { g_1}\def \gd { g_2}\def \gt { g_3}\def \gc { g_4}\def \gp { g_5}
\def \F {{\cal F}}
\def \del { \partial}
\def \t {\theta}
\def \p {\phi}
\def \ee {\epsilon}
\def \te {\tilde \epsilon}
\def \ps {\psi}

\newsection{ Superpotential for 5-brane on $S^2$  solution}
To complement the above discussion  of complex geometry
and  supersymmetry
of the \CV background, let us now demonstrate  
that,  like other $N=1$ supersymmetric ``D3-brane on conifold"
type IIB  solutions
\ci{kt,kst,pt}, 
 this 
solution (in its S-dual \CVD  form \ci{mn}) 
can be derived from a first order system which 
follows from a superpotential.
We shall also show that 
there exists a general ansatz  for the background fields 
and the 1-d  action for the radial evolution
of which  all of the solutions of \ci{kt,kst,pt} and \ci{mn}
are  special cases. That ansatz should be useful
for finding  more general 
solutions  that ``interpolate" between the
``D3 + wrapped D5 on  conifolds"  and ``D5 on $S^2$"
 backgrounds. 

\subsection{Interpolating ansatz for 3-branes on conifold and 
\CVD }

The ansatz  for the type IIB supergravity
 fields we shall choose below is motivated  by the special
 cases studied in \ci{kt,kst,pt,mn}.
 We shall consider the 10-d metrics of topology $\bR^{1,3}\times \bR \times S^2
  \times S^3$ with non-trivial functions depending only on
the radial direction. The conifold metrics have $SU(2)_{L 1}
 \times SU(2)_{L 2}$ symmetry. The \CV metric in \rf{dila} does not have 
 the  $Z_2$ symmetry between the two
2-spheres, so we are going to relax it, but 
we will insist on $SU(2)_{L 1} \times U(1)_{L2}$ symmetry.
We  shall use  the 
 Einstein frame metric and embed the S-dual 
 of the \CV background, i.e. the 3-form  in \rf{dila}
 will  appear as special case of the R-R 3-form
  $H_{RR} \equiv  F_3$.

\def \non {\nonumber}
\def \KK {{\cal K}}
\def \K {{\rm K}}

As we will be dealing with 6-spaces of $\bR \times S^2 \times S^3$ 
topology and  metrics having $SU(2) \times U(1)$ isometry, 
 let us define the  relevant 1-forms:
$\{e_1,e_2\}$  will  correspond to  the first $S^2$ (with coordinates 
$\t_1,\p_1$),  and 
$\{\ee_1,\ee_2,\ee_3\}$ will be the left-invariant forms on  $S^3$
with Euler angle coordinates $\ps,\t_2,\p_2$ (cf. 
\rf{minu},\rf{basi},\rf{noww}):
\bea e_1\equiv d\theta_1  \ ,  \ \ \ \  
e_2\equiv  - \sin\theta_1 d\phi_1  \ , \non \\
\ee_1\equiv \sin\psi\sin\theta_2 d\phi_2+\cos\psi d\theta_2
\ , \ \ \ \ \ \ \  
\ee_2  
\equiv   \cos\psi\sin\theta_2 d\phi_2 - \sin\psi d\theta_2
\ ,  \non \\
\te_3 = \ee_3 + \cos\theta_1 d\phi_1
 \equiv (d\psi + \cos\theta_2 d\phi_2) + \cos\theta_1 d\phi_1 
\ , \\
d \ee_i = - {1\ov 2}\ee_{ijk}  \ee_j \we \ee_k   \ . \non
\eea
Let us  also introduce\foot{To
 compare to the notation used in \ci{kst} (KS) and \ci{mn} (MN) 
note that:
 $e_1= (e_2)_{\rm KS} = (A_1/a)_{\rm MN} ,  \ \ 
e_2 =(e_1)_{\rm KS}= (A_2/a)_{\rm MN} , 
$ $
\ee_1
= (e_4)_{\rm KS} = (w_1)_{\rm MN},  $
$
\ee_2  
= (e_3)_{\rm KS} = (w_2)_{\rm MN} ,  $
$
\te_3
= (e_5)_{\rm KS}= (w_3 - A_3)_{\rm MN},$ 
and $\te_c = (w_c - A_c)_{\rm MN}$. 
The forms used in KS in our notation are:
$
g^1 = - {\ee_2-  e_2 \over\sqrt 2},\ 
g^2 = - {\ee_1  - e_1 \over\sqrt 2} , \ 
g^3 = {\ee_2 + e_2 \over\sqrt 2} ,\     
g^4 = {\ee_1 + e_1 \over\sqrt 2},\  
g^5 = \te_3 .
$ Note also  that $\phi_1 = -(\phi)_{\rm MN}$.}  
\bea
\te_1 \equiv  \ee_1 - a(u)  e_1\ , \ \ \ \  \ \ \  \ \ \ 
\te_2 \equiv  \ee_2 - a(u)  e_2\ ,
 \eea 
 where $a$ is a function of  the radial coordinate
 here  denoted  as $u$.
 
 Our  ansatz for the (Einstein-frame) 
 10-d metric  which includes 
  the metrics in \ci{kt,kst,pt} and \ci{mn} 
   as special cases is ($m=0,1,2,3$) 
\beq
ds^2_E =  e^{2 p-x} (e^{2A}  dx_m dx_m + du^2)  + ds_5^2 \ , 
\eeq 
\bea
ds_5^2 =  e^{x+g}  ( e_1^2 + e_2^2)     
      +   e^{x-g}  ( \te_1^2 + \te_2^2)  + e^{-6p -  x} \te_3^2
\la{dii} \ 
\eea
\bea
=\  ( e^{x+g} + a^2 e^{x-g} )  ( e_1^2 + e_2^2)  
+     
e^{x-g} [(   \ee_1^2 + \ee_2^2)  - 2 a ( \ee_1 e_1 + \ee_2 e_2) ] 
+ e^{-6p -  x} \te_3^2\ ,  \label{mmm}
\eea
where $p,x,A, g,a$ are functions of $u$ only.
The  function $a $ thus 
multiplies 
the ``off-diagonal"
term $\ee_1 e_1 + \ee_2 e_2$.
The $Z_2$ symmetry between  the two spheres   is broken unless
$e^{x+g} + a^2 e^{x-g} = e^{x-g}$, i.e. $e^{2g} = 1 -a^2$  
(note that $ \ee_1^2 + \ee_2^2 \to   e_1^2 + e_2^2$
under $(\t_1,\p_1) \to (\t_2,\p_2)$).
In the singular and resolved   conifold  cases $a=0$ \ci{kt,pt}.

To specialize  the above general  ansatz to the 
 case corresponding 
to the deformed conifold solution of \ci{kst} 
one is to  relate $g$ and $a$, replacing them 
by a single new function $y(u$) 
 as follows (see \ci{pt}):\foot{The case of deformed conifold \ci{min,oht,kst}  corresponds to 
$a = - (\cosh \tau)^{-1}$,
  where $\tau$ is related to $u$ so that 
the conifold metric takes the form
$ds^2_6 = \ha  \ee^{4/3} \KK \big[ (3 \KK^3)^{-1} ( d \tau^2  + \te_3^2) 
+ \ha  \sinh^2 \tau (\cosh \tau)^{-1} ( e_1^2 + e_2^2) 
+ \ha  \cosh \tau ( \te_1^2 + \te_2^2) \big] ,  
$ where $\KK^3= ( \ha \sinh 2 \tau - \tau )/\sinh^3 \tau$.}
   \beq
   e^{-g}=  \cosh y \  , \ \ \ \ \ \ \ \ \ \ 
    a=  \tanh y\ . \la{ttt} \eeq
In the case of the  D5-brane version of \CV solution 
\rf{mert}--\rf{dila} \ci{mn}  
\bea
A={ 2\ov 3} ( g + \P) \ , \ \ \ \  \ x= g+ \ha  \P    
 \ , \ \ \ \ \ \   \ \ \
    p= - {1\ov 6} ( g+ \P) \ ,\non\\ 
     a= { 2 u (\sinh 2 u)^{-1} }\ , \ \ \  \ \ 
dr = e^{ - {2\ov 3 }( g+\P) } du \ , \non\\
 e^{-2 \P} = 2 e^g  (\sinh 2 u)^{-1} \ , \ \ \ \ \    \ \ \ 
 e^{2 g} = u \coth 2 u - { 1 \ov 4} (1+ a^2) \ . \la{maln}
 \eea
 Here we  use the 
 Einstein-frame metric
 related to \CV string-frame metric \rf{gaa} by 
 $ds^2_E= e^{\P/2} (ds^2)_{NS5_{S^2}}$,  where $\P$ now is
  the \CVD dilaton, i.e.
 it has the opposite sign 
 compared to the one in  \rf{dila}. 
To  match  the expressions in \rf{gaa}
one  needs  to make a  rescaling 
$e^{\P/2}\to 4 e^{\P/2}, \ 
 e^{2g} \to  { 1 \ov 4} e^{2g} ,  \ r \to { 1 \ov 2} r  $
 as here 
we did  not include $1/4$ factors in the
$S^3$ part of the  metric \rf{dii} (cf.  \rf{mert}).

In addition to the dilaton $\P(u)$ we shall 
assume that only the two  3-form strengths 
and the 5-form 
field strengths of type IIB supergravity 
are non-zero.
Our  choice for these NS-NS and R-R  forms will be
$$
B_2  =   h_1(u)  (\ee_1 \we \ee_2 + e_1 \we e_2)+ 
\chi(u)  (-\ee_1 \we \ee_2 + e_1 \we e_2)
  + h_2(u)  (\ee_1 \we e_2 -  \ee_2 \we e_1 ) 
 \ , $$ \bea
  H_3 = d B_2 &=&  h_2(u)  \te_3 \we  (\ee_1 \we e_1 + \ee_2 \we e_2 )   
+  du \we \big[ h'_1(u)  (\ee_1 \we \ee_2 + e_1 \we e_2)
\non \\ &+&\ 
\chi'(u)  (-\ee_1 \we \ee_2 + e_1 \we e_2)
  + h'_2(u)  (\ee_1 \we e_2 -  \ee_2 \we e_1 )\big] 
 \ ,\la{hhh}   \eea \bea
F_3 = P  \te_3\we \big[  \ee_1 \we \ee_2 +  e_1 \we e_2 
-  b(u)  (\ee_1 \we e_2 - \ee_2 \we e_1) \big] \non 
\\
+  \ du \we \big[ b'(u) (\ee_1 \we e_1 + \ee_2 \we e_2) \big]
\ \la{ftri} , \eea\bea 
F_5 = {\cal F}_5  +  {\cal F}^*_5\ , 
\ \ \ \ \  \ \ \ 
 {\cal F}_5 = \K(u) e_1 \we e_2 \we \ee_1 \we \ee_2 \we \ee_3
\ . \la{fffff}
\eea
Here $h_1,h_2,\chi, b,\K$ are  functions of the radial direction 
$u$ only (primes denote derivatives over $u$) 
 and $P$ is a constant.
 It is straightforward to check that the backgrounds 
 in \ci{kst,pt} and \ci{mn} are special cases of this one.
Note that, in general, the ``off-diagonality" functions $a$ and $b$ 
in the metric \rf{mmm} and  in the R-R 3-form \rf{ftri} 
are different  (they are equal for 
the \CV solution).
 
It is useful to 
write the forms \rf{hhh},\rf{ftri},\rf{fffff}  
in the basis $\{e_a, \td \ep_i\}$ 
in which the metric \rf{dii} is diagonal:
\bea 
H_3   =  h_2   \te_3 \we 
 (\te_1 \we e_1 + \te_2 \we e_2 )   
+  du \we \bigg[ (h'_1-\chi') \te_1 \we \te_2\non\\
 + \  [h_1'(1+a^2) + 2h_2' a +\chi'(1-a^2) ]
e_1 \we e_2
+ (a h_1' + h'_2-a\chi')   (\te_1 \we e_2 -  \te_2 \we e_1 )\bigg] 
\ , \eea
\bea
F_3 =  P\bigg[ \te_3 \we \big[ \te_1 \we \te_2 
+  (a^2- 2 a b + 1)   e_1 \we e_2 
\non\\
+   \  (a-b)  (\te_1 \we e_2 - \te_2 \we e_1) \big]  
+ \ du \we \big[ b' (\te_1 \we e_1 + 
\te_2 \we e_2) \big] \bigg]\ , 
\eea \beq
{\cal F}_5 = \K(u) e_1 \we e_2 \we \te_1 \we \te_2 \we \te_3
\ . 
\eeq
To obtain the  1-d action that leads to the 
equations of motion for 
all the unknown functions 
of $u$ (i.e. $p,x,A, g,a$     and    $\P,h_1,h_2,\chi, b,\K$)
one may follow the discussion
in \ci{kt,pt}  and use the equation 
 for $F_5$  in combination 
with the  relevant parts of the 
type IIB supergravity action 
\bea  S =  { 1 \ov 4}   \int d^{10} x  \sqrt {g_E}  
\bigg[ R - \ha  (\del \P)^2  - \tv  e^{-\P}  H^2_3 -
\tv e^{\P}  F^2_3  - {1\ov 4\cdot  5!} F^2_5 \bigg] + {\rm
CS-term}\  . 
  \eea
The  effective 1-d action reproducing
the  equations of motion restricted to the above ansatz
has the following general structure
\begin{equation}
 S_1=   \int du
\ e^{4 A} (3A'^2 + L) 
=  \int du
\ e^{4 A} \bigg[ 3 A'^2
- \ha G_{ab}(\varphi)  \varphi'^a  \varphi'^b -
 V(\varphi)\bigg]
 \ ,  \la{vvv}  \eeq
 where $\varphi^a$ stand for all unknown functions of $u$.
 This action  should 
 be supplemented with   the ``zero-energy" constraint
 $
 3 A'^2
- \ha G_{ab}(\varphi)  \varphi'^a  \varphi'^b +
 V(\varphi)= 0.$
 Explicitly,   using  the metric in (\ref{mmm})  one obtains 
\beq
 { 1 \ov 4}\int d^9 x  \sqrt {g _{E}} \ R \ \to \ \ 
 e^{4A} ( 3 A'^2 + L_{gr})  \ , \eeq
 \beq
 L_{gr}  = 
  -  \ha  x'^2 - { 1 \ov 4} g'^2 - 3 p'^2  
 - { 1 \ov 4 } e^{-2 g} a'^2    - V_{gr} \ , \label{iyu}
\eeq
 \beq\label{uiy}
V_{gr} = -\ha e^{ 2 p - 2 x} [e^g   + (1+ a^2) e^{- g}]
 +   { 1 \ov 8} e^{- 4 p - 4 x} [  e^{ 2 g}   +  (a^2 -1)^2  e^{ -2 g}
  + 2 a^2 ] 
   + { 1 \ov 4} a^2 e^{- 2g + 8p } \ . 
  \eeq
The ``matter" part of the 1-d action is found to be  
$$
L_m = - { 1 \ov 8}  \bigg[  \P'^2 
   +     e^{-\P-2 x } 
   \bigg(   e^{  2g  } (h_1'-\chi')^2 
   $$
   $$
   +  \  e^{ - 2g  }  [ (1+ a^2) h_1' 
   + 2 a h_2'+(1-a^2)\chi']^2 
   + 2 (a h_1' + h_2'- a\chi')^2   +   2 e^{ 8p  } h_2^2 \bigg) 
   $$
   $$  +\ 
      P^2 e^{ \P - 2 x } \bigg( e^{ 8p } [ e^{ 2g }   
   + e^{- 2g } ( a^2 - 2 ab + 1)^2  +    
     2(a-b)^2  ]  + 2  b'^2 \bigg)  $$
\beq\label{maat}
    +  \   e^{8p - 4x }  [Q +  2P(h_1 + b  h_2)]^2  \bigg] \ ,  
   \eeq
   where we used that the type IIB supergravity 
   equation for $F_5$ implies  that 
   \beq
   \K(u) = Q +  2P[h_1(u)  + b(u)   h_2(u)] \ ,\ \   \ \ \ \ \ \ 
   Q=\const \ . \la{kek} \eeq  
Notice that $\chi$ enters the  1-d action \rf{iyu},\rf{maat} 
only through its derivative, 
so that it can be eliminated  using its equation of
 motion\foot{The derivation  and analysis 
 of the next two equations was done in collaboration  
 with  S. Frolov.}
$$
 e^{  2g  } (h_1'-\chi') 
   +   e^{ - 2g  } (a^2-1) [ (1+ a^2) h_1' + 2 a h_2'+(1-a^2)\chi']
   $$
   \beq
   + \ 2 a (a h_1' + h_2'- a\chi')=0 \ . \la{conn}
   \eeq
Then we get 
$$
L_m = - { 1 \ov 8}  \bigg[  \P'^2 
   +     e^{-\P-2 x } 
   \bigg(   2h'^2_2 
   +   4e^{ - 2g  }  (  h_1' + a h_2')^2 
   $$
   $$
   -\ 4[e^{2g}+(1-a^2)^2e^{-2g}+2a^2]^{-1}
   \big[ e^{-2g}(1-a^2)( h_1' + a h_2')-a h'_2\big]^2 
    +   2 e^{ 8p  } h_2^2 \bigg) 
   $$
   $$  +\ 
    P^2  e^{ \P - 2 x } \bigg( e^{ 8p } [ e^{ 2g }   
   + e^{- 2g } ( a^2 - 2 ab + 1)^2  +    
     2(a-b)^2  ]  + 2  b'^2 \bigg)  $$
 \beq\label{meaat}
    +  \   e^{8p - 4x } [   Q + 2P(  h_1 + b  h_2)]^2  \bigg] \ . 
 \eeq
 The cases considered in \ci{kt,kst,pt} and \ci{mn} 
 are consistent truncations of this system
 (in particular, they are  special solutions of 
 the resulting system  of equations):
 
 (i) The   fractional 3-brane on {\it singular }
 conifold solution \ci{kt} corresponds to 
 $a=b=\chi=h_2=g=0,\ \P=\const$  (it is a special case of the deformed 
 and resolved conifold backgrounds below).
 
 (ii) The action corresponding to the 3-brane 
 on {\it resolved} conifold 
solution  \ci{pt} 
is found for  $a=b=h_2=0,\ \P=\const$. 
In the notation of \ci{pt} 
\beq
h_1 = \ha (f_1 -f_2)\
, \ \ \ \ \ \ \ 
\chi= \ha( f_1 + f_2)\ , \ \ \ \  \ \ g=y\ . \la{reee}
\eeq
 
(iii) The 1-d action   corresponding to the 3-brane on 
{\it deformed}  conifold case \ci{kst}  is found \ci{pt} 
once one 
relates $a$ and $g$ according to    (\ref{ttt})
 and also chooses
   $\P=\const$  and     $\chi=0$
which is then a solution of \rf{conn}.
The relation to the functions used in \ci{kst,pt} is 
\beq
h_1 = \ha (f+k)\  , \  \ \ \ \ \ 
 h_2 = \ha (k-f) \  , \ \ \ \ \ 
b =  P^{-1} F  - 1 \ , \ \ \ \  \  
\ \ \ 
  a^2 = 1-e^{2g} = \tanh^2 y \ .   \la{deee}
\eeq

(iv) The  case of the $D5_{S^2}$  background \ci{mn} 
 is $h_1=h_2=\chi=0, \  b=a$. To satisfy the  equations 
 for $a$ and $b$  under  the constraint $a=b$ 
 one is also to 
 impose the following relations  
 \beq\label{tre}
   \P= - 6 p -  g \  , \ \ \ \  \ \ \ \ \
   x= g + \ha  \P  = \ha g - 3 p    
 \ , 
 \eeq
which lead  precisely to the \CVD
 ansatz \rf{maln}.\foot{Again, to relate 
 the variables  here to the  functions  
 in  \rf{dila}
we need to change the sign of the dilaton 
and  to rescale 
$e^{2g} \to  { 1 \ov 4} e^{2g}$.}

\subsection{Superpotentials for special cases}
 In general,   
   the existence of a superpotential 
 means that $V$ in \rf{vvv}
 can be represented in the form
\begin{equation}\la{super}
V ={  {1\over 8}} G^{ab} {\partial W\over \partial \varphi^a}
{\partial W\over \partial \varphi^b} - { {1\over 3}} W^2
\ .
\end{equation}
In this case the 2-nd order equations following from \rf{vvv}
and  the ``zero-energy"constraint
are satisfied on the solutions of the 1-st order system
 \begin{equation}
\varphi'^a = { \ha } G^{ab} { \partial  W
 \over \partial \varphi^b}
\ ,
\ \ \ \ \ \ \ \
A' = - {{ 1 \over 3}} W (\varphi) \ .\la{flo}
\end{equation}
In all of the  four cases discussed above the 1-d system 
is such that it admits  a simple  superpotential $W$.

Indeed, 
in the resolved  conifold case
\rf{reee}
we have 
 $\varphi^a=(x,y,p,\Phi,f_1,f_2)$ 
and the 1-d  Lagrangian  \rf{iyu},\rf{uiy},\rf{maat} is 
\beq
L= -\ha x'^2-  { 1 \ov 4} 
y'^2- 3 p'^2 -  { 1\ov 8}  \bigg[
\Phi'^2  +    e^{-\Phi- 2x}
 ( e^{ - 2y}  f'^2_1 + e^{ 2y} f'^2_2) \bigg]  - V  \ , \la{kin}
 \eeq
 \beq
V = { 1 \ov 4}  e^{-4p-4x}\cosh 2y - e^{2p-2x}\cosh y+
 { {1\over 8}} e^{8p}  \bigg(  2 P^2 e^{\Phi-2x}\cosh 2y
+   e^{-4x} [Q + P(f_1 -f_2)]^2 \bigg)\ . \la{pot}
\eeq
The associated  superpotential is \ci{pt} 
\beq
W =     
      e^{ -2 p - 2 x} \cosh y  +  e^{   4 p  }  +
 \ha e^{4p - 2x } [Q+ P (f_1 -f_2)]  \ .  
 \  \la{sii}
 \eeq
 The special ``symmetric" 
 solution $y=0$  (implying also $f_1=-f_2$, i.e.
   $\chi=0$) 
 corresponds to the  singular  conifold case \ci{kt}.
 
   In the deformed conifold case  \rf{deee} 
   \ci{kst}  
   we have 
    $\varphi^a=(x,y,p,\Phi,f,k,F)$
    and   the 1-d action  \rf{iyu},\rf{uiy},\rf{maat}  becomes
 \beq
L= - \ha x'^2-  { 1 \ov 4} 
y'^2 -3 p'^2  -  { 1\ov 8}  \bigg[
\Phi'^2  +    e^{-\Phi- 2x}
 ( e^{ - 2y}  f'^2 + e^{ 2y} k'^2) + 
 2  e^{\Phi- 2x} F'^2  \bigg]     - V  \ , \la{kiin}
 \eeq
   $$
V=  { 1 \ov 4}  e^{-4p-4x}- e^{2p-2x}\cosh y 
 +{ 1 \ov 4}   e^{8p}\sinh^2 y  
 + 
 { {1\over 8}} e^{8p}  \bigg[ \ha  e^{-\Phi-2x} (f-k)^2 
 $$ \beq
 + \  e^{\Phi-2x}  [ e^{  -2y}  F^2  + e^{ 2y} (2P-F)^2]
+  e^{-4x}   [Q +  k  F + f (2P-F)]^2  
  \bigg]\ ,  \la{pote}
\eeq 
The corresponding  superpotential  \ci{pt} 
  has the form similar to 
\rf{sii} 
   \beq W 
    = \ e^{-2 p-2x}  +  e^{4p} \cosh y 
   +   \ha  e^{4p - 2x } [ Q +  k  F+ f(2P-F)]
    \ .  \la{wewa}
   \eeq
Note that, as in \ci{kt}, $W$
 in these two cases 
happens to be dilaton-independent (so that $\P=\const$ is a
solution)
and is simply 
a sum of a gravitational and matter parts.

In the \CVD  case  the 1-d action takes the form 
(using  $h_1=h_2=\chi=0, \ b=a$ and \rf{tre})
\bea
L &=&   -   { 1 \ov 2} g'^2 
 - { { 1 \ov 2}  }  e^{-2g} a'^2 -  12  p'^2    - V  \ , \la{mree} \\
V&=&  {  1 \ov 4  } e^{ 8 p } \bigg[
 (a^2 -1)^2 e^{-4 g} -2 e^{-2 g} -  1 \bigg] \ . 
 \eea
The   superpotential for this system 
is quite different from \rf{sii},\rf{wewa}\foot{This  
difference should have  to do with  the fact that the 
 dilaton is now  non-constant 
 and is  mixed  with the gravitational functions
 via  \rf{tre}. Note also that the kinetic term metric 
 is no longer flat but is  that of $AdS_2 \times \bR$.}
 \beq\label{ssuu}
 W=  e^{ 4 p } \sqrt{(a^2 -1)^2 e^{-4 g} 
 + 2 (a^2 + 1) e^{-2 g}+ 1 } \ . \la{mnmn} 
 \eeq
 The first  order system \rf{flo}  following  from 
 \rf{mree},\rf{ssuu}  is then solved by \rf{maln}.
 
It would be very interesting to find other 
special cases of the above general 
ansatz that also admit superpotentials
(and thus are likely to lead to new 
 supersymmetric solutions).
  There must be at least one generalization of 
  the  wrapped 5-brane
   solution: it should be possible to extend 
   the background of \ci{mn} 
  away from the 5-brane throat region.\foot{To describe
   this case  one needs to 
   modify the original  ansatz of \ci{cv}
   to include an extra scalar
  corresponding to the radius of $S^3$.}

\section*{Acknowledgments}
We are grateful  to S. Frolov,  
L. Pando Zayas  and  I. Klebanov
for 
useful
discussions. A.T.  would like to 
thank S. Frolov  for a collaboration on  attempts
to find new examples of backgrounds
admitting superpotentials  generalizing 
 the ones discussed in Section 5.
This work is
partially supported by SPG grant
PPA/G/S/1998/00613.
G.P. is supported by a
University Research Fellowship from
the
Royal Society. 
 The work of A.A.T.  is 
supported in part by
the DOE grant DE-FG02-91R-40690,
EC TMR grant ERBFMRX-CT96-0045 
and 
INTAS project 991590.
We would like also to thank  CERN 
Theory Division 
for hospitality while
 part of this work was done. 
  

\def\appendix#1{
  \addtocounter{section}{1}
  \setcounter{equation}{0}
  \renewcommand{\thesection}{\Alph{section}}
  \section*{Appendix \thesection\protect\indent \parbox[t]{11.65cm}
  {#1} }
  \addcontentsline{toc}{section}{Appendix \thesection\ \ \ #1}
  }

\setcounter{section}{0}
\setcounter{subsection}{0}

\appendix{ Solution of $J^2=-1$ condition for \CV}

\def \zz {{\rm z}} 

Here we   solve the equations \rf{oo},\rf{ooo},\rf{oooo}
arising from the condition $J^2=-1$ on the complex structure.
Let us  determine
 $z_i{}^a$  in terms of  3-vectors $\l_i$ and $\m_i$ and 
 2-vector $w_a$. 
 Solving the second equation in (\ref{ooo})
 we get (up or down position of indices  does not 
 matter as 
 we use Euclidean signature)
 \beq
 z^a_i = \zz^a_i  + (P^2 - \m^2)^{-1} \bigg[ 
 (P \l_i - P^{-1}  \m \cdot \l\  \m_i )\  \epsilon_{ab} w_b  +
 \epsilon_{ijk} \l_j \m_k\  w_a \bigg]\ , 
 \la{gg}
 \eeq 
 where $\zz^a_i$ 
 is solution of homogeneous equation  in (\ref{ooo}).
 Then from (\ref{oo}) we find that
 \beq
 \l^2=1-  w^2 \ , \  \ \ \ \ \m^2= P^2+ w^2 \ , \ \ \ \ \ 
\m \cdot \l\equiv  \m_i \l_i  =  P \ . 
\la{res}
\eeq
Direct check shows that 
the first equation 
 in (\ref{ooo})  and (\ref{oooo})
are satisfied identically.
 $\zz^a_i$ is the general solution if either $\l^i$ or $w_a$ are
 equal to zero. Writing $\zz^a_i = ( x_i, y_i)$ we find
 from the second equation  in (\ref{ooo})
 \beq\la{vv}
 P x_i = \epsilon_{ijk} \m_j   y_k\ , \ \ \ \ \ \ \ \ \ \ \  
  P y_i = - \epsilon_{ijk} \m_j   x_k\ ,
 \eeq
 so that for $P\not=0$ 
 \beq
 y \cdot \m\  \m_i = (\m^2-P^2) y_i \ , \ \ \ \ \ \ \  
 x \cdot \m\  \m_i = -(\m^2-P^2) x_i \ .  \eeq
{}From  (\ref{oo}) we learn that for $w_a=0$ we have
$\l_i$ as  a unit 3-vector orthogonal to $x_i$ and $y_i$, 
and  parallel to  $\m_i$.  From  (\ref{oooo})
$x^2_i = y^2_i= 1-P^2 $. 
Thus we can choose 
$\m^2= P^2$,\   $\l_i = (0,0, 1)$,\  $\m_i= (0,0,-P)$, 
and from (\ref{vv})
\beq x_i= (X,0,0)\ , \ \ \ \ \ 
 y_i= (0,-X,0)\ , \ \  \ \ \ X^2 = 1-P^2  \ . \eeq 
This is  the ``radial" solution \rf{aa},\rf{aa},\rf{squa}, 
with the asymptotic case \rf{asa},\rf{assa} 
being $X=0$.

For $w\not=0$  we can also write  (\ref{gg}) as 
\beq
 z^a_i =  \zz^a_i- w^{-2}  \bigg[ 
 (P \l_i -  \m_i )  \epsilon_{ab} w_b  +
 \epsilon_{ijk} \l_j \m_k  w_a \bigg] \ . 
 \la{ggg}
 \eeq 

 \appendix{ The Calabi metrics}

Let us consider 
a  metric is on the resolved $\bZ_m$ singularity of $\bC^m$. The 
K\"ahler potential is 
\beq K=(r^{2m}+1)^{1\over m}+{1\over m} 
\sum_{j=0}^{m-1} \zeta^j \log \big[(r^{2m}+1)^{1\over m}-\zeta^j\big] 
\ , \eeq 
where $\zeta=e^{2\pi i\over m}$ and $r^2=\delta_{i\bar j}z^i 
\bar z^{\bar j}$ ($i,j=1,\dots,m$). The coordinates $z^i$ are
 those of $\bC^m$.
Observe that these metrics have a $U(m)$ isometry group.

There may be a generalization of these Calabi spaces \ci{who}
 to include
torsion. For this consider the ansatz for the metric
\bea
ds^2=\big[\delta_{i\bar j} h(r^2) +
\bar z_i  z_{\bar j} k(r^2)\big]
dz^i d\bar z^{\bar j}\ ,
\eea
where $\bar z_i=\delta_{i\bar j} \bar z^{\bar j}$ and $z_{\bar i}=\delta_{j\bar i}
z^j$.
It is well known that the associated 
torsion three-form $H$ can be
determined from the metric and the complex structure. 
In particular,
we have
\bea
H=
\big[\delta_{i \bar j} z_{\bar k} (h'-k)-
 \delta_{i \bar k} z_{\bar j} (h'-k)\big]
dz^i \wedge d \bar z^{\bar j}\wedge d\bar z^{\bar k}+ c.c.\ ,
\eea
where primes denote differentiation with respect to $r^2$.
It remains to show that $H$ is closed. This leads to the
condition that
\bea
2(\delta_{k \bar i}\delta_{j \bar\ell }
-\delta_{j \bar i}\delta_{k \bar\ell }) (h'-k)
\delta_{k\bar i} z_{\bar \ell} \bar z_j (h''-k')- ~~~~& &
\nonumber \\
\delta_{j\bar i} z_{\bar \ell}
\bar z_{k} (h''-k')-\delta_{k\bar \ell} z_{\bar i} \bar z_j (h''-k')
+\delta_{j\bar \ell} z_{\bar i} \bar z_k (h''-k')&=&0\ .
\eea
It turns out that if the dimension of the manifold is more than four,
then the  above equation has solutions provided that
\beq
h'-k=0\ ,
\eeq
which implies that the torsion vanishes. So the only solution
is that given by the Calabi metrics. In four dimensions, there
is a solution with non-vanishing torsion 
that is represented  by the
NS5-brane background.



\end{document}